\documentclass[pra,showpacs,onecolumn,superscriptaddress,amsmath,amssymb]{revtex4-1}
 
 \pdfoutput=1

\usepackage{graphicx}
\usepackage{color}
\usepackage{dcolumn,amsmath}
\usepackage[normalem]{ulem}
\usepackage{xfrac}

\usepackage{dcolumn}
\usepackage{bm}
\usepackage{amssymb}

\begin{document}

\title{Effects of Conical Intersections on Hyperfine Quenching of Hydroxyl OH in collision with an ultracold Sr atom}

\author{Ming Li}
\affiliation{Department of Physics, Temple University, Philadelphia, Pennsylvania 19122, USA}
\author{Jacek K{\l}os}
\affiliation{Department of Physics, Temple University, Philadelphia, Pennsylvania 19122, USA}
\affiliation{Department of Physics, Joint Quantum Institute, University of Maryland, College Park, MD, 20742, USA}
\author{Alexander Petrov}
\affiliation{Department of Physics, Temple University, Philadelphia, Pennsylvania 19122, USA}
\affiliation{NRC, Kurchatov Institute PNPI, Gatchina, 188300, and Division of Quantum Mechanics, Saint Petersburg State University, St. Petersburg, 199034, Russia}
\author{Hui Li}
\affiliation{Department of Physics, Temple University, Philadelphia, Pennsylvania 19122, USA}
\author{Svetlana Kotochigova}
\email{skotoch@temple.edu}
\affiliation{Department of Physics, Temple University, Philadelphia, Pennsylvania 19122, USA}

\date{\today}

\begin{abstract}

The effect of conical intersections (CIs) on  electronic relaxation, transitions from excited states to ground 
states, is well studied, but their influence on hyperfine quenching  in a reactant molecule is not known.  Here, 
we report on ultracold collision dynamics of the hydroxyl free-radical OH with Sr atoms leading to
quenching of  OH hyperfine states. Our quantum-mechanical calculations of this process reveal that 
quenching is efficient due to anomalous molecular dynamics in the vicinity of the conical intersection at 
collinear geometry. We observe wide  scattering resonance features 
in both elastic and inelastic rate coefficients at collision energies below  $k_{\rm B}\times10$ mK.
They are identified as either $p$- or $d$-wave shape resonances. 
We also describe the electronic potentials relevant for these non-reactive collisions, their diabatization procedure, as 
well as the non-adiabatic coupling between the diabatic potentials near the CIs.  
\end{abstract}

\maketitle

\section{Introduction}

A diverse list of promising applications for ultracold molecular processes governed by quantum 
mechanics exists. This includes creating new types of sensors, advancing quantum information science,
simulation of complex exotic materials, performing precision spectroscopy to test the Standard Model of 
particle physics, and, excitingly, the promise of quantum control of chemical reactions as each molecule can 
be prepared in a unique ro-vibrational quantum state. In all these applications the ability to control the internal 
and external degrees of freedom of the cold molecular species is essential.  
At high temperatures, the state population distribution in molecular gasses is often spread over many 
quantum states, obscuring  the key role of quantum effects. 
Therefore, over the past three decades much attention of the scientific community has been directed to 
developing techniques to cool molecules to cold ($< 1$ K) and ultracold ($<1$ mK) temperatures as well as study their collective properties in the ultracold domain. 

The coldest di-atomic molecules  are now produced by magneto- and photo-association from laser-cooled 
alkali-metal atoms \cite{Ye2017}. In fact, the associated dimers are prepared in their lowest ro-vibrational level 
and have translational temperatures below 1~$\mu$K. 
Despite this success, the association method only provides access to a limited range of molecular species. 
There  exist many molecules, both di-atomic and polyatomic, that can not be easily formed from ultracold 
atoms.  One such molecule is the hydroxyl radical OH as oxygen atoms has not been yet laser cooled.  This 
molecule has attracted attention due to promising applications in precision measurement and quantum 
computation \cite{Lev2006, Hudson2006, Kozlov2009, Fast2018}. In addition, OH is an important molecule in 
understanding the behavior of interstellar media, astrophysical research \cite{Green1981,Bochinski2003,Yusef2003},
and atmospheric and climate science \cite{Rex2014}. 

Evaporative cooling of hydroxyl, OH molecules loaded from Stark decelerators has so far been the method of choice to create molecular gasses with a few mK temperature \cite{Stuhl2012}. Further cooling of these molecules to temperatures below 1 mK is desirable. One of the promising direct cooling techniques, proposed in Ref.~\cite{Soldan2004,Hudson2009}, is the sympathetic cooling of  molecules in thermal contact with  laser-cooled neutral atoms. There have been  numerous proposals to achieve sympathetic cooling using different atom-molecule pairs. See,  for example, Refs.~\cite{Wallis2009, Wallis2011}.  However, for various combinations of atoms and molecules inelastic collisions are projected to occur more frequently than elastic collisions leading to overwhelming trap losses. A favorable elastic-to-inelastic ratio has been predicted for collisions of OH  with atomic hydrogen with initial OH temperatures  around 250 mK \cite{Hutson2013}.

Sympathetic cooling of the internal states of neutral molecules with laser-cooled neutral atoms is far more challenging.  Trap depths for 
neutral molecules are small and the energy release from  vibrational, rotational, and even hyperfine relaxation rapidly decreases the number 
of cold molecules.  On the other hand molecular ions, trapped in deep Penning or Paul traps, can survive sympathetic cooling of the internal states.
The method was first demonstrated for the BaCl$^+$ molecular ion colliding with ultracold calcium \cite{Nature2013}
and further analyzed in Ref.~\cite{Stoecklin2016}. Collisions of this molecule ion with a polarizable ultracold atom encourages the two to thermalize via vibrational relaxation and elastic momentum-changing collisions. 

\begin{figure*}[t]	
\includegraphics[width=1\columnwidth,trim=0 0 20 0,clip]{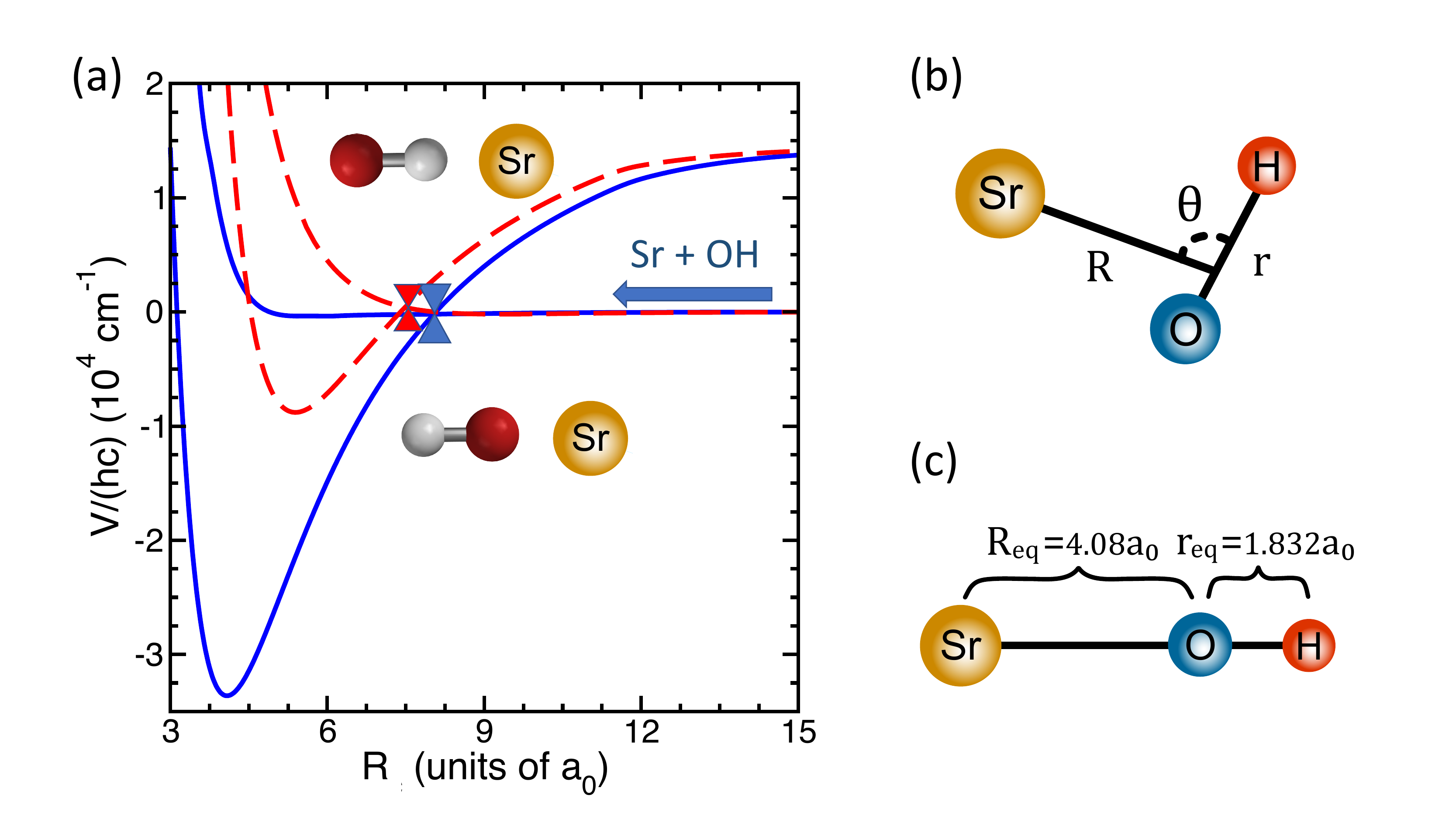} \hspace*{1cm}
\caption {Relevant two-dimensional potential energy surfaces of non-reactive Sr+OH($^2\Pi$) as functions of Jacobi coordinates $R$. 
(a) Potential energies of the  1\,$^2\!A^\prime({\rm X}^2\Sigma^+)$ and 4\,$^2\!A^\prime({\rm F}^2\Pi)$ states as functions of $R$ at $\theta = 180^\circ$ (blue solid curves) and $\theta = 0^\circ$ (dashed-red curves), respectively.  Red and blue cones indicate conical intersections between X and F curves. They are located at $R=7.4 a_0$  and $8.1 a_0$, respectively. The arrow marks the entrance channel. (b)  A schematic depiction of the Jacobi coordinates.
(c) Jacobi coordinates of the SrOH molecule in optimized geometry.}
\label{fig:engfile}
\end{figure*}

Here, we examine the possibility of sympathetic cooling of external and internal degrees of freedom of OH molecules in their electronic X$^2\Pi_{3/2}$ ground state due to collisions with ultracold ground-state Sr atoms. We have developed a state-of-the-art quantum coupled-channel description of the OH+Sr system, thereby, allowing the calculations of collisional (elastic) and quenching (inelastic) rate coefficients.
Our earlier study of the  collisional complex SrOH  \cite{Li2019} has proven that symmetry-required conical intersections (CIs) exist between its ground  and  excited $^2\!A^\prime$ adiabatic electronic potential energy surfaces (PESs). 
Conical intersections are sets of degeneracy points between adiabatic PESs and are common features in the electronic structure of polyatomic molecules \cite{Domcke2004,Matsika2011}.  They play an important role in ultrafast radiation-less transitions from  excited to ground electronic states found in photochemistry, molecular spectroscopy, and quantum reactive scattering 
\cite{Domcke2012,Hoffman2000,Kendrick2015,Kendrick2018}. Yet, there remain many  fields of physics and chemistry, where the effects of CIs on the collisional dynamics of molecules is not fully understood. Additional studies  are required. We focus on the role of CIs in the hyperfine quenching of OH molecules in collisions with ultracold Sr atoms. It is our goal to answer the natural question whether a conical intersection in the collisional complex can influence the quenching dynamics 
of the free radical.   

\section{Results}
\subsection{Potential Surfaces involved in collisional dynamics}

The non-reactive collision physics  of $^{88}$Sr($^1{\rm S}$) colliding with the tightly-bound ${v=0,\, J=3/2}$ ro-vibrational ground-state of $^{16}$O$^1$H(X$^2\Pi_{3/2}$) molecules is most conveniently described in three-dimensional Jacobi coordinates or vectors $\boldsymbol{R}$ and $\boldsymbol{r}$, where $R=|\boldsymbol{R}|$ is the separation between Sr and the center of mass of OH and $r=|\boldsymbol{r}|$ is the separation between O and H. It is also useful to define angle  $\theta$ between $\boldsymbol{R}$ and $\boldsymbol{r}$. A schematic of the Jacobi coordinates is shown in Fig.~\ref{fig:engfile}(b). 

\begin{figure*}
	\includegraphics[width=1\columnwidth,trim=0 0 0 0,clip]{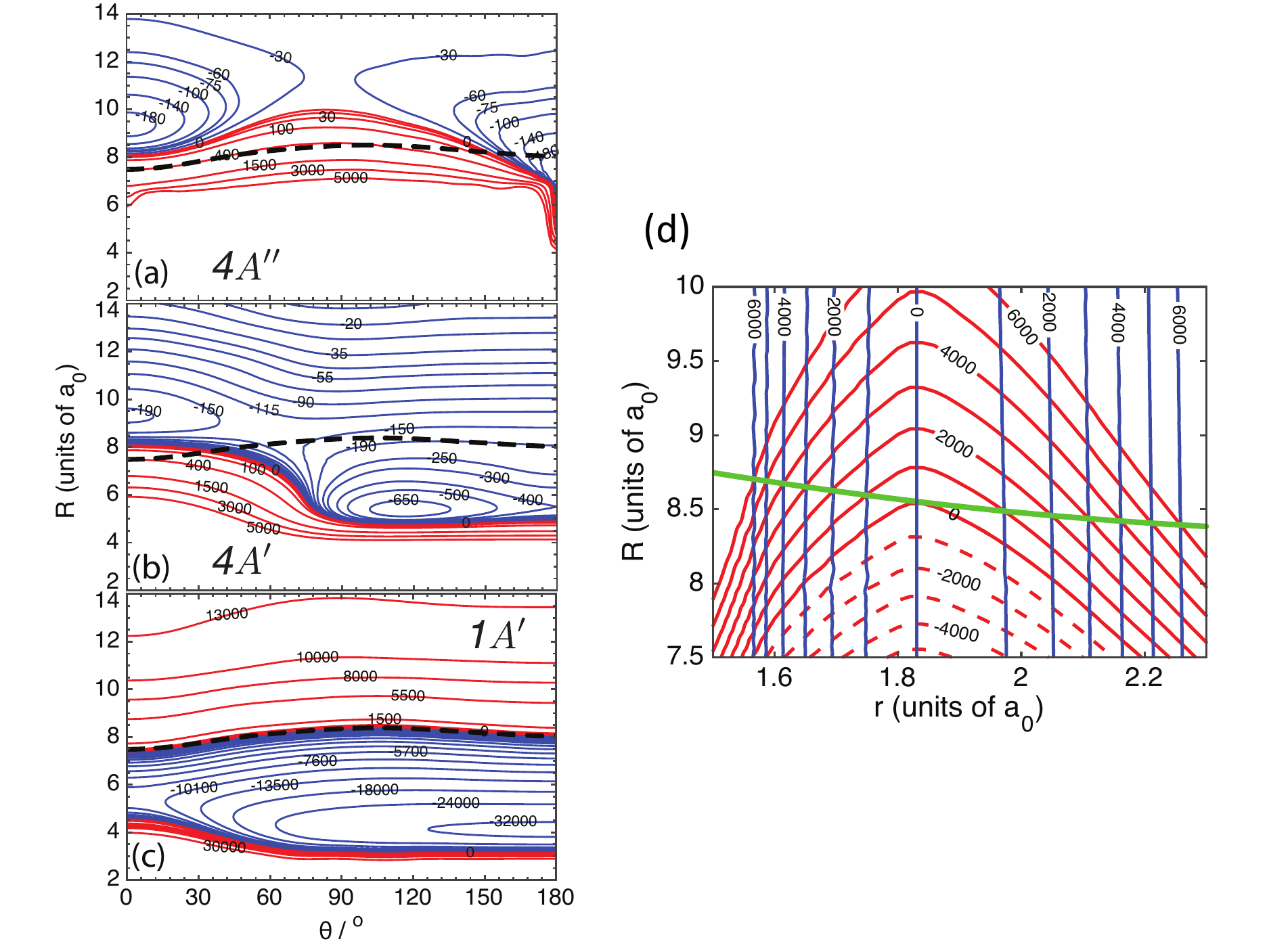}
	\caption{ Contour plots of {\it diabatized} non-relativistic Sr+OH PESs.
	(a), (b), and (c) Contour plots of $4\,^2\!A^{\prime \prime}({\rm F}^2\Pi)$, $4\,^2\!A^{\prime}({\rm F}^2\Pi)$, and $1\,^2\!A^{\prime}({\rm X}^2\Sigma^+)$ states, respectively, as functions of Jacobi coordinates $R$ and $\theta$. Red and blue contours are labeled by potential energies  in the units of cm$^{-1}$.  The dashed  black curve in each panel corresponds to locations where energies of the $1\,^2\!A^{\prime}$ and $4\,^2\!A^{\prime}$ states are equal. (d) Contour plot of the $1\,^2\!A^{\prime}({\rm X}^2\Sigma^+)$ (red curves) and $4\,^2\!A^{\prime}({\rm F}^2\Pi)$ (nearly-vertical blue curves) states as  functions of Jacobi coordinates $R$ and $r$  for the collinear Sr-O-H arrangement  with ${\theta=180^\circ}$. The seam of conical intersections is  shown by the green curve. Contours are labeled by energies are in units of cm$^{-1}$ and spaced by 1000 cm$^{-1}$ for both potentials. The zero of energy is at the Sr($^1$S)+OH(X$^2\Pi$) dissociation limit. The OH separation is fixed at ${r=1.832 a_0}$.}
	\label{fig:pescon}
\end{figure*}

The relevant electronic structure of the tri-atomic system is well characterized by three non-relativisitic {\it diabatic} 
electronic potential surfaces: two shallow nearly-degenerate at large $R$ van-der-Waals-bonded potentials 
labeled by $4\,^2\!A^{\prime}({\rm F}^2\Pi)$ and $4\,^2\!A^{\prime \prime}({\rm F}^2\Pi)$ that dissociate to ground-state Sr($^1{\rm S}$) and OH($^2\Pi$) and one deep ionically-bound potential labeled by $1\,^2\!A^{\prime}({\rm X}^2\Sigma^+)$. These surfaces are shown in Figs.~\ref{fig:engfile} and \ref{fig:pescon}. The potential surface of the ionically-bound $1\,^2\!A^{\prime}({\rm X}^2\Sigma^+)$ state dissociates to an electronically-excited state of the Sr atom and the ground state OH(X$^2\Pi$) molecule. In our notation of states trimer symmetries $^2\!A^{\prime}$ and $^2\!A^{\prime \prime}$ are further specified by $^2\Lambda^\pm$ labels in parenthesis. These describe the $C_{\infty{\rm v}}$ symmetries of the electronic wavefunctions in co-linear geometries. It is convenient to denote trimer potentials by $C_{\infty{\rm v}}$ labels as the equilibrium geometry of SrOH is linear with O in the center. In fact, the equilibrium separation between O and H is close to that of the X$^2\Pi$ ground-state potential of the OH dimer
and the vibrational energy in the OH stretch is only 3\,\% larger than in the OH dimer. For our calculations it then also suffices to determine the three PESs for only a small range of OH separations $r$ around the OH dimer equilibrium separation of $1.832 a_0$, where $a_0$ is the Bohr radius.

Finally, trimer states $m^2\!A^{\prime}$ are labeled by integer $m=1,2,\cdots$ and to a lesser extent by characters X and F in the parenthesis. 
The value $m$ is follows the energy ordering of  $^2\!A^{\prime}$ PESs near the trimer equilibrium geometry. As the diabatic $4\,^2\!A^{\prime}$ and 
$4\,^2\!A^{\prime \prime}$ states are nearly degenerate at linear optimized geometry, the $^2\!A^{\prime \prime}$ is also denoted by $m=4$ even though it is not the fourth state with this symmetry. The excited $2\,^2\!A^{\prime}$ and $3\,^2\,A^{\prime}$ electronic states have been omitted in our description of the Sr+OH collision. These potentials cross the $4\,^2\!A^{\prime}({\rm F}^2\Pi)$  potential at much smaller separations $R$ and the effects of their couplings are expected to be smaller.

Figure~\ref{fig:engfile}(a) shows the two $^2\!A'$ electronic potentials at $\theta=0^\circ$ and $180^\circ$, co-linear 
$C_{\infty{\rm v}}$ geometries, as functions of  separation $R$ with $r=1.832 a_0$. The deeper and shallower
of the two potentials correspond to the  $1\,^2\!A'({\rm X}^2\Sigma^+)$ and  $4\,^2\!A'({\rm F}^2\Pi)$ states, respectively.
In these co-linear geometries  the two states have conical intersections near $8a_0$. The energy of the CI for $\theta=180^\circ$ is lower than that of our entrance channel Sr($^1$S)+OH(X$^2\Pi$). The corresponding energy for $\theta=0^\circ$ is higher and thus classically forbidden.
Figure \ref{fig:engfile}(b) shows two-dimensional cuts through the diabatic potential energy surfaces of the $1\,^2A'$, $4\,^2\!A'$, and 
$4\,^2\!A''$ electronic states as functions of radial separation $R$ and angle $\theta$.  
The $4\,^2\!A'$ and $4\,^2\!A''$ PESs are indistinguishable on the scale of the figure.  We observe that the ground state crosses the two excited state along a curve of almost constant $R$. Their coupling, to be described below, is only zero at $\theta=0^\circ$ and $180^\circ$,
the location of the CIs.

\begin{figure*}
         \includegraphics[width=1\columnwidth,trim=0 0 0 0,clip]{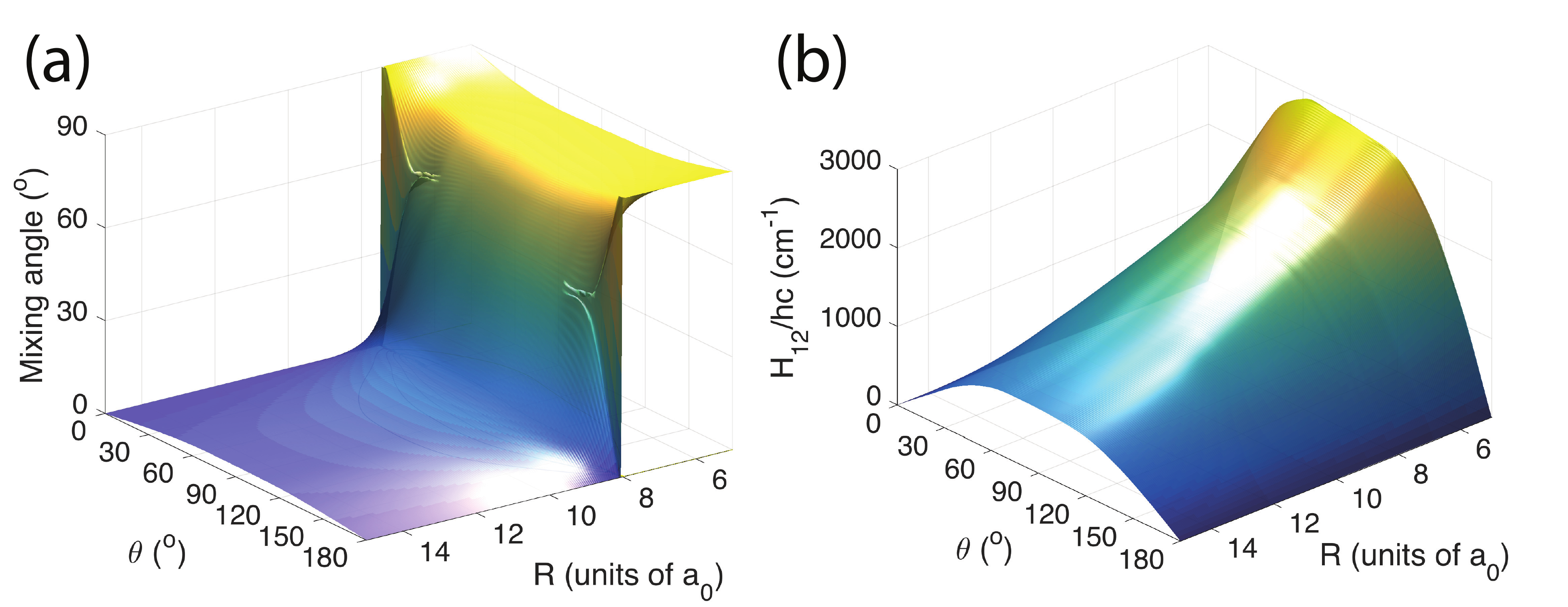} 
	\caption{Mixing angle $\beta(R,r,\theta)$ (panel a) and coupling potential $H_{12}(R,r,\theta)$ (panel b) between the diabatic $1\,^2\!A'$ and $4\,^2\!A'$ states at ${r=1.832 a_0}$ as functions of Jacobi coordinates $R$ and $\theta$.}
\label{fig:H12}
\end{figure*}
We have determined  these diabatic potentials and their couplings from  {\it ab~initio} non-relativistic coupled-cluster electronic-structure calculation of the SrOH PESs with single, double, and perturbative triple excitations (CCSD(T)) and the equation-of-motion coupled cluster (EOM-CCSD(dT)) method described in detail in our previous study \cite{Li2019}. The procedures to diabatize the adiabatic potential energy surfaces, as determined by the non-relativistic electronic-structure calculations,  have also been described in this reference.

Figure~\ref{fig:pescon} shows a more quantitative view of the three diabatic potentials.
In particular, the difference between the shallow  potentials of the $4\,^2\!A^{\prime\prime}$ and $4\,^2\!A^{\prime}$ states 
is apparent in panels (a) and (b), respectively. In fact, their depth is less than $hc\times400$ cm$^{-1}$ and $hc\times700$ cm$^{-1}$, 
respectively.  In addition, the PES of the $4\,^2\!A^{\prime\prime}$ state is very shallow in the T-shape $\theta=90^\circ$ region, exhibiting 
a saddle point between the two collinear minima. The PES of $4\,^2\!A^{\prime}$  is more attractive, allowing a closer approach of the Sr atom 
towards OH in a skewed geometry near $\theta=120^\circ$. The $4\,^2\!A^{\prime}$ state has  two saddle points: one collinear with 
$\theta=180^\circ$ and one around $\theta=60^\circ$.  The $1\,^2\!A^{\prime}$ potential is very deep and shown in Fig.~\ref{fig:pescon}(c). 
In all panels we have indicated the curve where  the $1\,^2\!A^{\prime}$ and $4\,^2\!A^{\prime}$ states have the same energy.
On this curve conical intersections occur at $\theta=0^\circ$ and $180^{\circ}$. 

For heteronuclear tri-atomic molecules conical intersections are not isolated molecular geometries, but  form a one-dimensional {\it seam}.
The molecule can undergo nonadiabatic passage or transitions at any point of this seam. In SrOH the seam between the $1\,^2\!A^{\prime}$ and 
$4\,^2\!A^{\prime}$ potentials lies in the $(R,r)$ plane with angle $\theta$ at either $0^\circ$ or $180^\circ$. Figure~\ref{fig:pescon}(d)  shows
the  $1\,^2\!A^{\prime}$ and $4\,^2\!A^{\prime}$ potentials in this plane for $\theta=180^\circ$ as well as the seam located between  $R=8.3a_0$ and 
$8.7a_0$. The equal-energy contours of the  $4\,^2\!A^{\prime}$ state are nearly independent of $r$.
As this diabatic state dissociates to Sr($^1$S)+OH($^2\Pi$) for large $R$ and the OH($^2\Pi$) state has a zero-point energy of nearly 
$hc\times 2000$ cm$^{-1}$, only a limited region of $r$ around $1.832a_0$ is relevant. For the diabatic $1\,^2\!A'$ state the contours are 
curved in the $(R,r)$ plane with the largest $R$ value very close to the OH equilibrium separation. In fact, the curvature at the equilibrium 
separation is almost independent of contour especially for contours with energy less than that of the Sr($^1$S)+OH($^2\Pi$) limit.
Equivalently, except for an energy off set, the $1\,^2\!A^{\prime}$ potential as function of $r$ close to $1.832a_0$ is nearly independent of $R$.
In fact, as noted before the OH separation and zero-point energy of the OH stretch in the SrOH trimer at its equilibrium geometry 
(not shown in Fig.~\ref{fig:pescon}(d)) are close to that of the OH dimer.
We will then assume that in the collision OH vibrational motion in the $1\,^2\!A'$ state is  limited to a small range of $r$ around $1.832a_0$.

We require diabatic PESs for use in  coupled-channels calculations of the Sr+OH collision. Only adiabatic PESs, however, are available from electronic structure calculations, here provided by the MOLPRO program \cite{molpro2012}. In addition, we computed the mixing angles $\beta(R,r,\theta)$ for the adiabatic $^2\!A^{\prime}$ states by the DDR procedure within the MRCI method \cite{Li2019}. The corresponding adiabatic states and potentials were transformed into diabatic states and potentials described in the previous subsection, so that the problematic avoided crossings near and degenerate singular part at the CI seam are ``removed'' \cite{Abrol2002}. By construction diabatic states are independent of the Jacobi coordinates near the CI and coupled by coupling potentials $H_{1,2}(R,r,\theta)$ between diabatic states $1$ and $2$. Specifically, near the CI seam between the $1\,^2\!A^{\prime}$ and $4\,^2\!A^{\prime}$ states we first computed a $2\times2$ orthogonal matrix $\bm O(R,r,\theta)$, parameterized by mixing angle $\beta(R,r,\theta)$, to transfer between the adiabatic and diabatic states. From the orthogonal matrix the coupling function can be easily constructed. For details see Ref.~\cite{Li2019}.

Figure~\ref{fig:H12} shows the mixing angle $\beta(R,r,\theta)$ and coupling potential $H_{12}(R,r,\theta)$ between $1\,^2\!A^{\prime}$ and $4\,^2\!A^{\prime}$ diabatic states as a function of Jacobi coordinates $R$ and $\theta$. At the CI the mixing angle changes rapidly from 90$^\circ$ to 0$^\circ$. The value of the coupling function is zero at $\theta = 0^\circ$ and $180^\circ$ and has a maximal strength of approximately $3300$ cm$^{-1}$ near $\theta=90^\circ$ and $R=8.5a_0$. For large $R$ the coupling relatively fast decreasing.

\subsection {Dynamics in the vicinity of a conical intersection}

\begin{figure}	
\includegraphics[width=0.7\columnwidth,trim=15 0 0 0,clip]{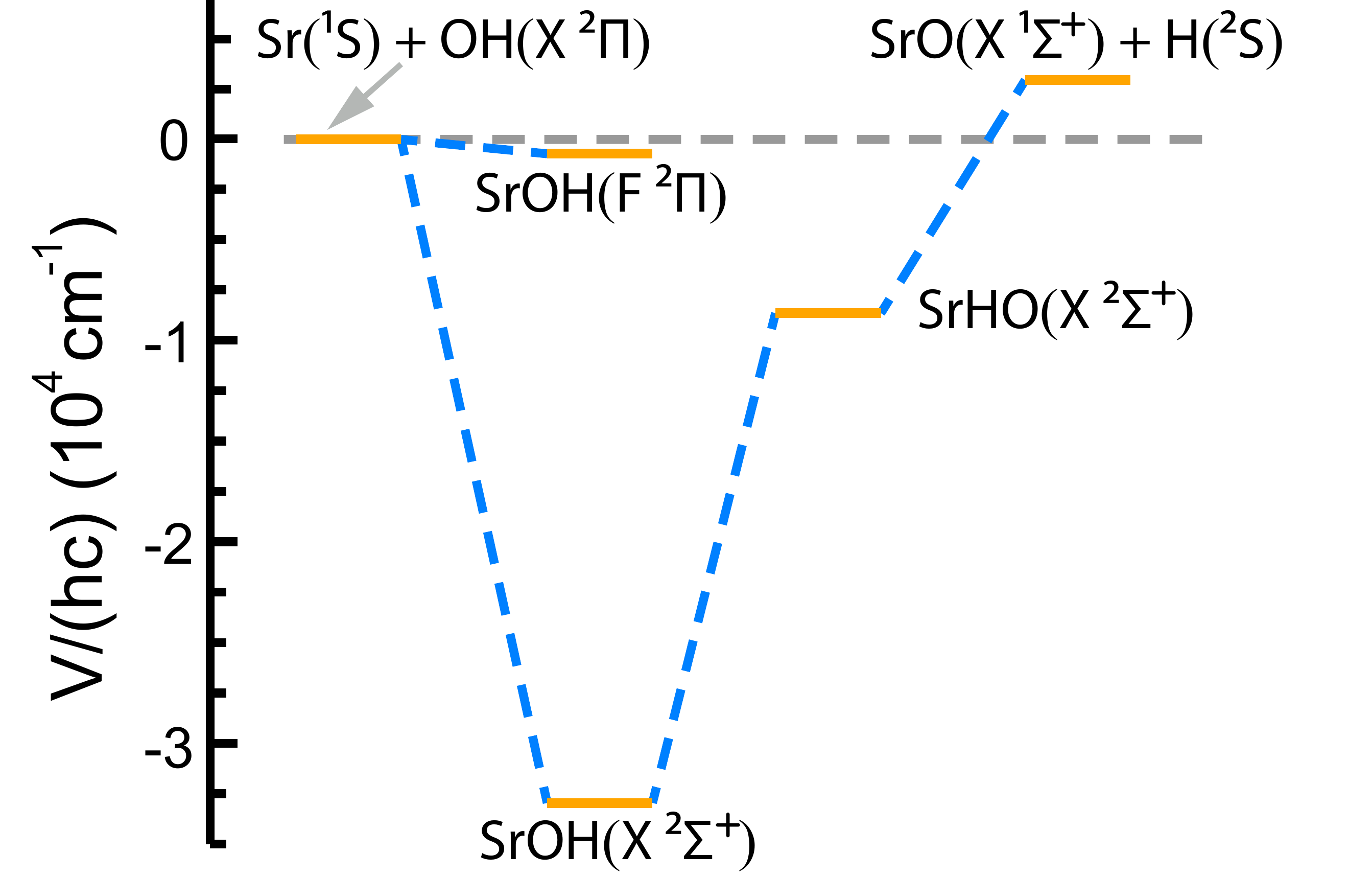}	
	\caption{Collision or reaction path of the Sr+OH system. The 
Sr($^1$S)+OH(X$^2\Pi$) entrance channel is shown on the left. Its energy is the zero of energy in the graph. The  SrO(X$^1\Sigma^+$)+H($^2$S) channel shown on the right  is endothermic by $hc\times2890$ cm$^{-1}$ and  includes the  difference in zero-point energy of OH and SrO.  
The three states with negative energies in between are trimer states with collinear geometries that are (local) minima of the $1\,^2\!A^{\prime}({\rm X}^2\Sigma^+)$ and $4\,^2\!A^{\prime}({\rm F}^2\Pi)$ electronic potentials as indicated. Electronic energies are taken from  {\it ab~initio} CCSD(T) calculations of Ref.~\cite{Li2019}. }
	\label{fig:diagram}
\end{figure}

To assess the collision of $^{88}$Sr($^1$S) with the rovibrational ground state of  $^{16}$O$^1$H(X$^2\Pi$)  it is
 convenient to first introduce  relevant energy correlation diagram in Fig.~\ref{fig:diagram}.
It illustrates the reaction pathway of our system, which highlights stationary points (minima or saddle points) in
the PESs. The pathway is based on  {\it ab-initio} calculations described in Ref.~\cite{Li2019}. 
The reaction to form SrO(X$^1\Sigma^+$)+H($^2$S)  is endothermic by $hc\times2890$ cm$^{-1}$ when accounting for the significant 
difference in zero point energy of OH and SrO.  The entrance-channel van-der-Waals interaction between Sr and OH creates a relatively
shallow $hc\times730$ cm$^{-1}$ deep well in the $4\,^2\!A'({\rm F}^2\Pi)$ collisional complex  that is non-adiabatically coupled to the  
$hc\times33\,600$ cm$^{-1}$ deep well in the $1\,^2\!A'({\rm X}^2\Sigma^+)$ collisional complex. 

\begin{figure}
	\includegraphics[width=0.6\columnwidth,trim=0 50 20 15,clip]{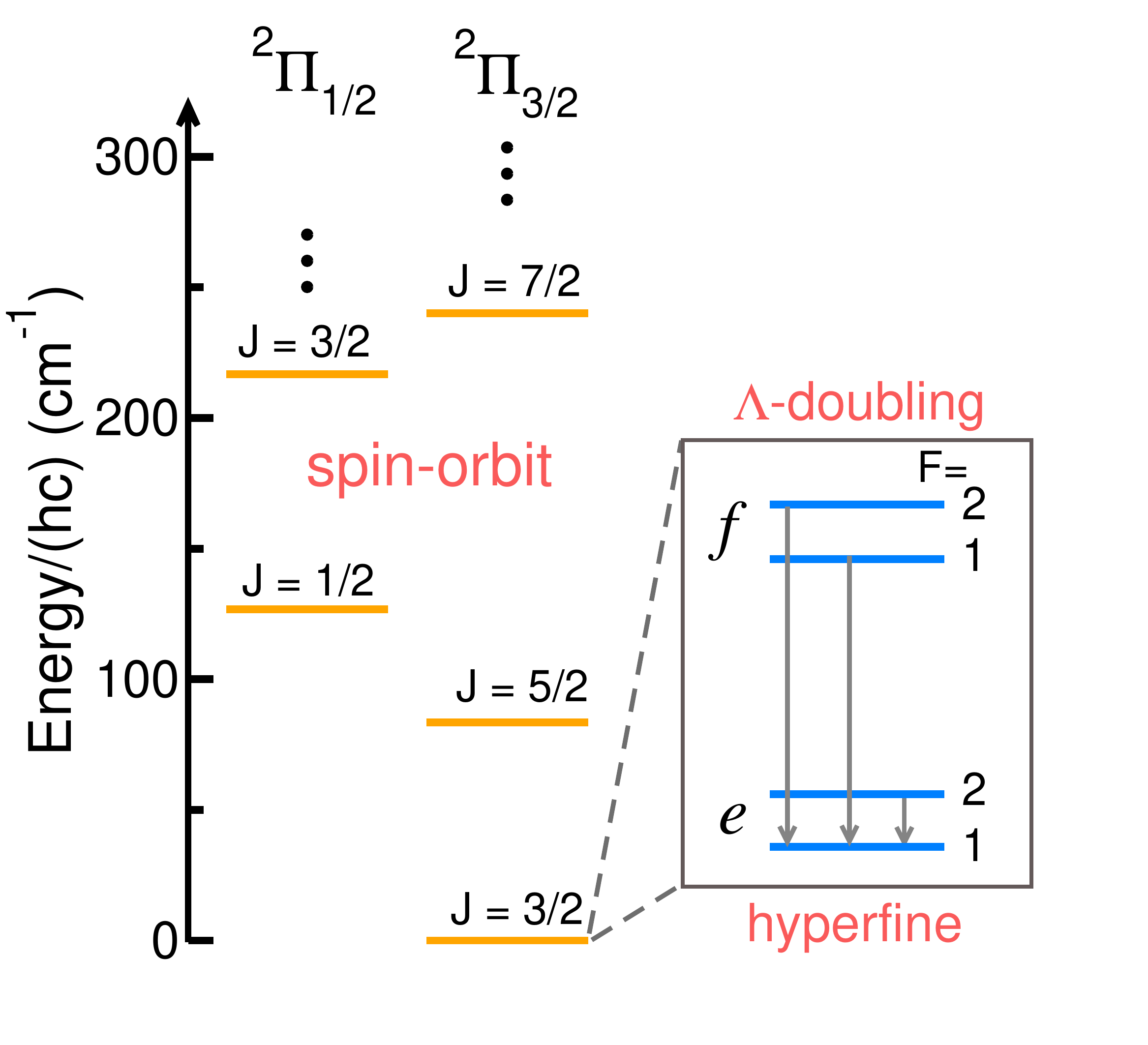}
	\caption{Energies of the ${v=0}$ vibrational level of the X$^2\Pi_{\Omega}$ state of $^{16}$O$^1$H. The left-hand side of the figure shows the  ${\Omega=1/2}$ and 3/2 spin-orbit splitting as well as a  rotational  progression in $J$. A blowup of the less than $hc\times0.1$ cm$^{-1}$ $\Lambda$-doubling (labeled by $e$ and $f$) and hyperfine splittings ($F=1$ and 2) of the energetically-lowest ${J=3/2}$ level are shown on the right.  Hyperfine splittings are due to coupling between the nuclear spin and electronic motion. 
	Arrows indicate the relevant collision-induced transitions between hyperfine components within $^{16}$O$^1$H.
	Their transition energies are given in the text.
	The zero of energy is at the energetically-lowest hyperfine state.}
	\label{fig:scheme}
\end{figure}

The relevant energy diagram of the $v=0$ vibrational state of $^{16}$O$^1$H(X$^2\Pi$)  is shown in Fig.~\ref{fig:scheme} based on 
spectroscopic data of Ref.~\cite{Maeda2015}. The energy spacing between the $v=0$ and $v=1$ vibrational levels of this state 
is about $hc\times 3570$ cm$^{-1}$ (not shown). The $v=0$ level has $^2\Pi_{\Omega=1/2}$ and $^2\Pi_{\Omega=3/2}$ spin-orbit
components separated by just over $hc\times 100$ cm$^{-1}$. The energetically-lowest rotational level is a $J=3/2$ state.  
The nuclear spin $\boldsymbol{I}$ of the hydrogen atom in $^{16}$OH leads to hyperfine splittings of the ground state rotational levels into 
components labeled by quantum number $F$, where $\boldsymbol{F}=\boldsymbol{J}+\boldsymbol{I}$ is the total angular momentum of $^{16}$OH. Finally, the ${\Omega=3/2}$ state is  split into opposite parity states $e$ and $f$ due to a weak non-adiabatic coupling, called $\Lambda$-doubling.    
The transition energies between hyperfine and $\Lambda$-doubling states of the $v=0,J=3/2,\Omega=3/2$ level of $^{16}$OH(X$^2\Pi$) relevant 
for this paper are 
\begin{eqnarray}
\nonumber
\Delta E({|}f; F=2 \rangle-{|}e; F^\prime=1 \rangle)/hc& = &0.057~{\rm cm}^{-1}, \\
\nonumber
\Delta E({|}f; F=1 \rangle-{|}e; F^\prime=1 \rangle)/hc& = &0.055~{\rm cm}^{-1}, \\
\nonumber
\Delta E({|}e; F=2\rangle-{|}e; F^\prime=1 \rangle)/hc& = &0.0018~{\rm cm}^{-1},
\end{eqnarray}
where in kets $|{e/f}; F \rangle$ and $|{e/f}; F' \rangle$ we omitted labels for the electronic and ro-vibrational state of $^{16}$OH for clarity.

We have developed a coupled-channels model to calculate atom-dimer quenching rate coefficients for $^{88}$Sr+$^{16}$O$^1$H collision energies 
below $hc\times 10^{-2}$ cm$^{-1}$ or $k_{\rm B}\times10$ mK, where $k_{\rm B}$ is the Boltzmann constant, in a rigid rotor 
approximation for OH.  That is, under the assumption that the OH stretch in SrOH is not excited in the Sr+OH collision.
The model includes the three diabatic non-relativistic $^2\!A'$ and $^2\!A''$ trimer potential  surfaces and their couplings as well as  
the spin-orbit, rotational, $\Lambda$-doubling, and hyperfine interactions in the OH dimer.
The vibrational wavefunction of the OH stretch is that of the $v=0$ eigenstate of the OH dimer for both the $^2\!A'$ and 
$^2\!A''$ symmetries.
As  discussed in the previous section this assumption is mainly justified by the observation that near the equilibrium geometry
of the $1\,^2\!A'({\rm X}^2\Sigma^+)$ state the vibrational energy of the OH stretch is close to that in the dimer.
Secondly, the depth of the potentials in the $4\,^2\!A'({\rm F}^2\Pi)$ and $4\,^2\!A''({\rm F}^2\Pi)$ states is 
less than the ground vibrational spacing of the OH(X$^2\Pi$) dimer.

Hence, we expand that the six-dimensional scattering wavefunction as a superposition of products of spherical 
harmonics with relative orbital angular momentum $\boldsymbol{\ell}$ in the orientation $\hat{\boldsymbol{R}}$ of coordinate $\boldsymbol{R}$ and rigid-rotor $v=0$ ro-vibrational states for OH in $\boldsymbol{r}$, such that the total trimer angular momentum 
$\boldsymbol{F}_{\rm tot}=\boldsymbol{\ell}+\boldsymbol{F}$ and parity are conserved. For this system negative (positive) parity
correspond to states or channels with odd(even) value for partial wave $\ell$.
Details about the basis set  are given in Methods.

The expansion leads to coupled Schr\"odinger equations in separation $R$ with potential matrix element $U_{ij}(R)$ between basis
set elements $i$ and $j$.  We  have numerically solved the differential equations and for $R\to\infty$ construct rate coefficients from
the solutions. The coupled-channels calculations are computationally demanding as a large number of OH rotational channels $J$ need 
to be included.  The  $^2\!A'$ and $^2\!A''$ potentials have strong angular anisotropies that couple many OH spin-orbit, rotational and hyperfine states. 
In the remainder of this paper we will study  elastic and inelastic rate coefficients of  Sr and OH in  states 
$| f;F=1,2 \rangle$ or $|e; F=1,2 \rangle$  states, defined in Fig.~\ref{fig:scheme}. 
In addition, we  intend to show how the conical intersection influences the outcome of the collision. 
This is achieved by switching the CI ``off'' by simply setting $H_{12}(R,r,\theta)=0$. 

\begin{figure}
	\includegraphics[width=0.7\columnwidth,trim=0 0 0 0,clip]{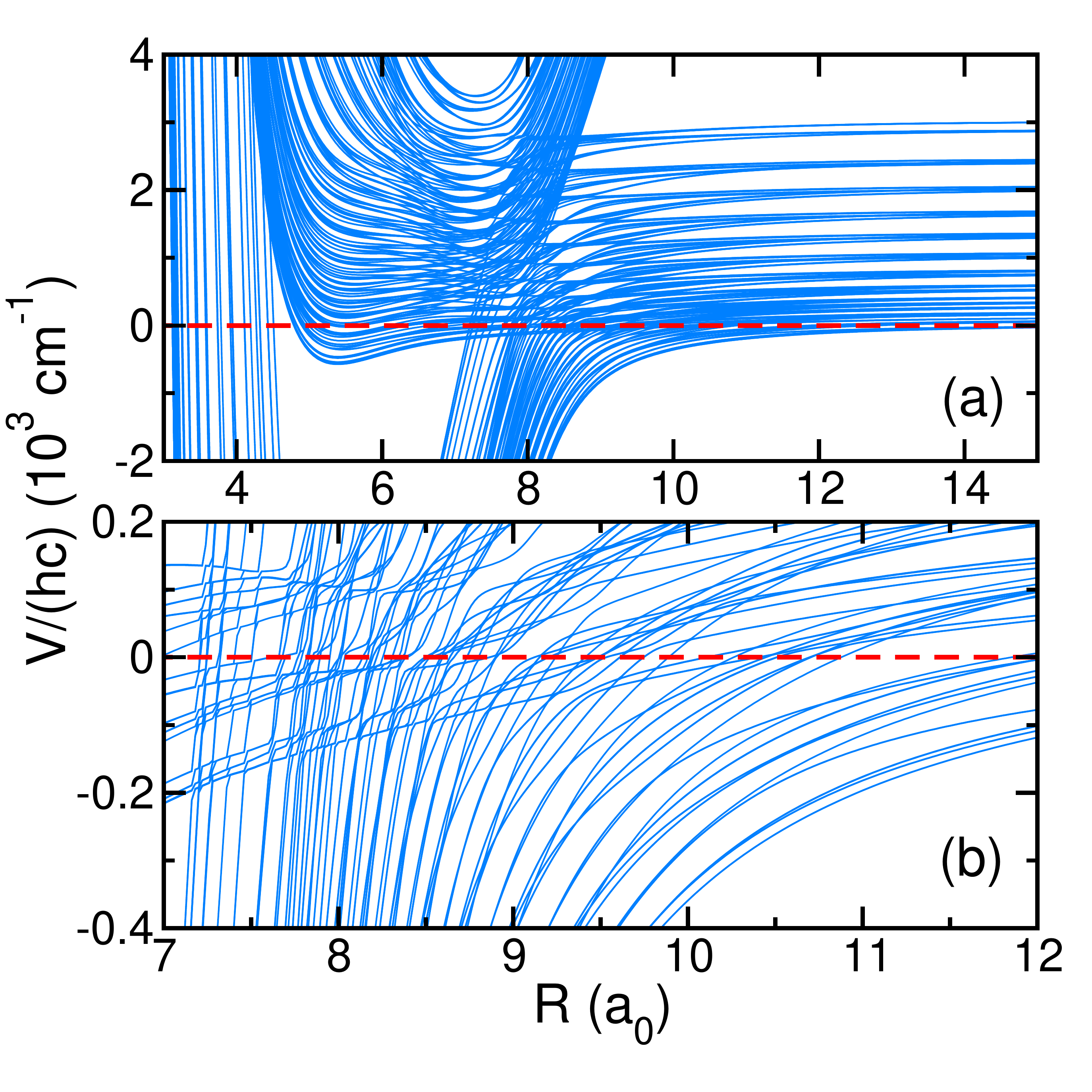}
	\caption{Adiabatic potential energies with $F_{\rm tot}=2$ and even parity of the $^{88}$Sr with 
	$v=0,J=3/2$ $^{16}$OH($^{2}\Pi_{3/2}$) collision as functions of $R$. Panels (a) and (b) show different energy scales of the
	same potentials. OH rotational states up to $J=25/2$ are included. The zero of energy is at the energetically-lowest ${F=2}$ hyperfine state of OH.}
	\label{fig:adiabats}
\end{figure}

Figure~\ref{fig:adiabats} shows the $F_{\rm tot}=2$ and even parity eigenvalues of matrix $U_{ij}(R)$ as functions of $R$ at 
two energy and length scales. We have included the energetically-lowest 12 rotational states of the $v=0$ X$^2\Pi_{1/2}$ and 
$v=0$ X$^2\Pi_{3/2}$ states of $^{16}$OH. These eigenvalues are also called adiabatic potentials, but now in the sense that only motion in the radial $R$ direction is considered slow. Most importantly, we observe that the density of collisional channels is high with spacings 
much smaller than the coupling  strength $H_{12}(R,r,\theta)$ providing by the CI.  In fact, near $R=8a_0$ in both 
Figs.~\ref{fig:adiabats}(a) and (b) a ``bundle'' of nearly-vertical curves avoids, in a complicated manner,
a bundle of nearly-horizontal curves. The former bundle is due to the $1\,^2\!A^{\prime}({\rm X}^2\Sigma^+)$ state. The latter
is due to both $4\,^2\!A^{\prime}({\rm F}^2\Pi)$ and $4\,^2\!A''({\rm F}^2\Pi)$ states. Their avoided crossings are due to the CI.

\begin{figure*}
	\includegraphics[width=1\columnwidth,trim=0 0 0 0,clip]{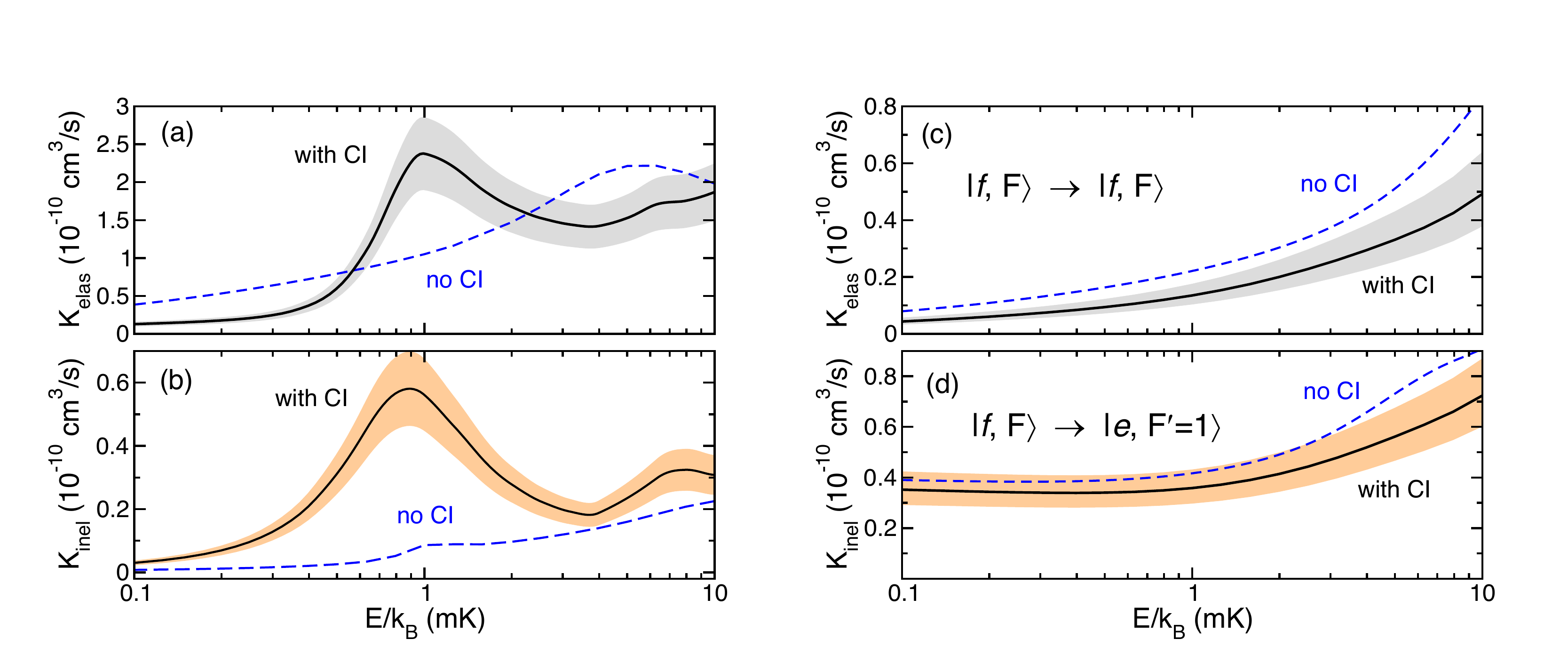}
	\caption{Collision energy dependence of the total elastic, panels (a) and (c), and inelastic, panels (b) and (d), rate coefficients for $^{88}$Sr+$^{16}$O$^1$H with $v=0,J=3/2$ $^{16}$OH(X$^2\Pi_{3/2}$). Solid and dashed lines indicate results of coupled-channels calculations with and without conical intersection between $1\,^2\!A^{\prime}$ and $4\,^2\!A^{\prime}$ potentials, respectively. OH $v=0$ rotational states up to $J=25/2$ are included. The grey and orange bands reflect our estimate of the one-standard deviation uncertainty of the rate coefficient due to our inability to include all rotational states of OH when the CI is included in the calculations. Panels (a) and (b) show total elastic and partial inelastic hyperfine quenching rate coefficients for OH in its $|e;F=2\rangle$ hyperfine state, whereas panels (c) and (d) show rate coefficients for OH in its ${|}f; F=1 \rangle$ and ${|}f; F=2 \rangle$ hyperfine states. Rate coefficients for ${|}f; F=1 \rangle$ and ${|}f; F=2 \rangle$ are indistinguishable  on the scale of the figure.}
\label{fig:rate2}
\end{figure*}

Figure~\ref{fig:rate2} shows the total elastic and hyperfine quenching inelastic rate coefficients obtained from coupled-channels calculations as a function of collision energy $E$ for OH in the hyperfine entrance channel  ${|}e;F=2 \rangle$, ${|}f;F=2 \rangle$, and ${|}f;F=1 \rangle$. Quenching leads to OH molecules in state  ${|}e;F^\prime=1 \rangle$.  For these calculations, we have included the energetically-lowest 12 rotational states of 
the $v=0$ X$^2\Pi_{1/2}$ and $v=0$ X$^2\Pi_{3/2}$ states of OH. Collisions with partial wave quantum numbers up to $\ell =8$  contribute to our entrance channels and collision energies $E/k_{\rm B}<10$ mK. Rate coefficients calculated when the coupling near the CI is switched on and off are shown.  

Figures ~\ref{fig:rate2}(a) and (b) show elastic and inelastic hyperfine quenching rate coefficients when OH is prepared in hyperfine entrance channel  ${|}e; F=2 \rangle \rightarrow {|}e; F^\prime=1\rangle$. Both rate coefficients have a resonance feature with a maximum value near $E=k_{\rm B}\times1$ mK but only when the coupling between the $1\,^2\!A^{\prime}$ and $4\,^2\!A^{\prime}$ states is switched on. 
This resonance is much weaker when the coupling is turned off. We observe that for both cases the elastic rate coefficients are about five time larger than the inelastic ones. Figures~\ref{fig:rate2}(c) and (d) show elastic and inelastic hyperfine quenching rate coefficients when OH is prepared in either hyperfine entrance channel  ${|}f; F=2 \rangle \rightarrow {|}e; F^\prime=1\rangle$ or ${|}f; F=1 \rangle \rightarrow {|}e; F^\prime=1 \rangle$. The rate coefficients are indistinguishable on the scale of the figure.

Analysis, based on studying the contributions from individual $F_{\rm tot}$ and parity channels,  has shown 
that the resonances that appear in Figures~\ref{fig:rate2}(a) and (b) are due to a shape resonance behind and tunneling through a ${\ell=1}$, $p$-wave centrifugal barrier leading to quantum enhanced scattering.  Similarly, the broader resonances near $E=k_{\rm B}\times8$ mK with CI coupling and near 
$E=k_{\rm B}\times5$ mK without CI coupling are found to be due to ${\ell=2}$, $d$-wave shape resonances.

\begin{figure}
	\includegraphics[width=0.6\columnwidth,trim=0 10 0 10,clip]{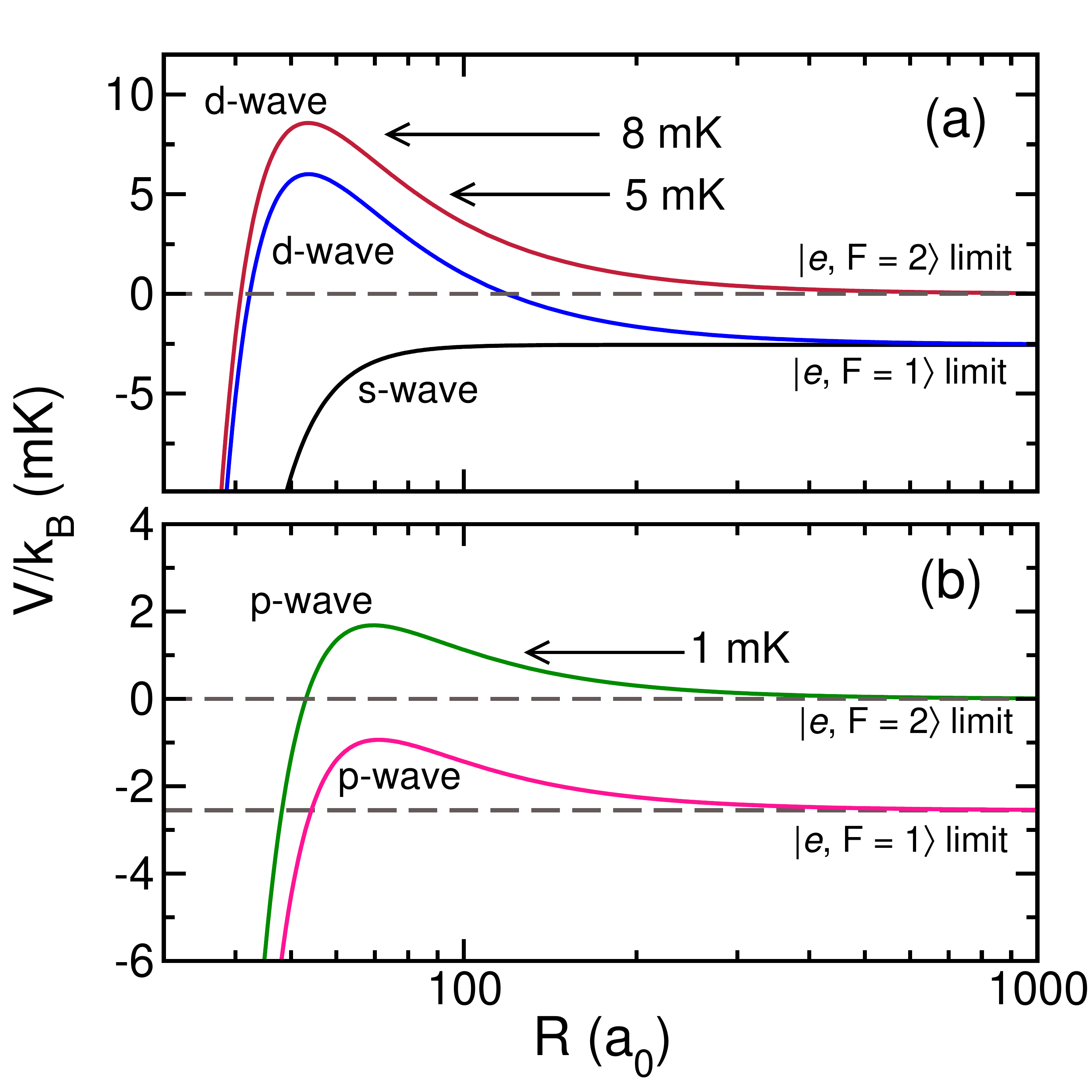}
	\caption{Centrifugal $s$, $p$, and $d$-wave barriers as functions of $R$ for the ${F_{\rm tot}=1}$ $^{88}$Sr+$^{16}$OH collision near the $^{16}$OH ${|}e; F=1 \rangle$ and ${|}e; F=2 \rangle$  hyperfine thresholds based on the adiabatic potentials in the coupled-channels calculations. Panels a) and b) show potentials for even and odd valued partial waves, respectively. The zero of energy is at the ${|}e; F=2 \rangle$ $^{16}$OH threshold. Arrows with collision energies $E/k_{\rm B}$ in mK indicate the locations of collisional resonances  in Fig.~\ref{fig:rate2}. }
	\label{fig:barrier}
\end{figure}

Figure~\ref{fig:barrier} explains the analysis of the resonances. The figure shows  the relevant ${F_{\rm tot}=1}$ adiabatic eigenvalues of matrix $U_{ij}(R)$  for large $R$ close to the OH ${|}e;F=1 \rangle$ and ${|}e;F=2 \rangle$ dissociation limits. Panels (a) and (b) show potentials for even and odd partial wave channels, respectively. On these energy and length scales centrifugal barriers for the  $p$ and $d$ wave channels are visible
with approximate barrier heights of about $k_{\rm B}\times 1$ mK and ${k_{\rm B}\times 10}$ mK, respectively.
Higher partial wave entrance channels have even higher barriers and do not significantly contribute to rate coefficients for collision energies 
below $k_{\rm B}\times 10$ mK. These barriers are not shown in the figure even though the corresponding channels are included in our 
coupled channels calculations. For our ${|}e;F=2 \rangle$ entrance channel and $F_{\rm tot}=1$ the $\ell=0$ state does not exist.
(The $s$-wave channel for the Sr+OH ${|}e;F=2 \rangle$ collision only occurs when $F_{\rm tot}=2$.)
The heights or tops of the $p$- and $d$-wave barriers are thus consistent with the two resonances when the coupling between 
$1\,^2\!A^{\prime}$ and $4\,^2\!A^{\prime}$ states is turned on. The location of the resonance at $E=k_{\rm B}\times5$ mK
when the coupling is turned off can only be explained by the $d$-wave centrifugal barrier for the
${|}e;F=1 \rangle$ exit channel, whose dissociation energy is about $k_{\rm B}\times 2.5$ mK lower in energy than
that of the ${|}e;F=2 \rangle$ entrance channel.

Finally, we have to note that with current computing capabilities first-principle calculations can not be fully converged with respect to 
the number of OH rotational states $J$ included in our coupled-channels calculations when 100\% of the coupling between the $1\,^2\!A^{\prime}$ and $4\,^2\!A^{\prime}$ states is turned on. These uncertainties are indicated in Fig.~\ref{fig:rate2} by colored bands and estimated from calculations of rate coefficients with different numbers of included OH rotational states. We have reached convergence when the coupling between the potential surfaces is turned off.  
Figure~\ref{fig:rate2}  also show that the inelastic rate coefficients obey the Wigner threshold law and approach a finite value in limit of zero collision energy. 

\section{Discussion}

We have created computational tools to compute elastic and quenching rate coefficients of the ultracold $^{88}$Sr+$^{16}$OH collision in the presence of nonadiabatic coupling between potential surfaces. These tools allowed us to treat the electronic structure and the nonadiabatic nuclear dynamics, dominated by conical intersections,  with quantum chemical and quantum dynamical methods. In particular, we have performed state-of-the-art calculations of non-reactive collisional dynamics of hydroxyl with Sr based on three electronic potential-energy surfaces using only a localized region of the tri-atomic nuclear coordinate space.

An important aspect of the investigation has been to answer the question whether conical intersections affect the quenching
of the hyperfine levels of the ultracold ground-state OH molecule. We therefore calculated collisional quenching rates under conditions where the conical intersection can be switched on and off. This was achieved by using diabatic electronic basis functions.
The results have shown that at low collision energy the conical intersection has important consequences for the collisional properties of OH.
Specifically, when OH is prepared in one of its energetically-lowest hyperfine states shape resonances can be observed, which
change their location or are absent when the coupling is turned off.

\section{Methods}

\noindent
{\bf Basis set}: The ultracold non-reactive $^{88}$Sr+$^{16}$OH scattering wave function $|\Psi\rangle$ is expanded in $^{16}$OH molecular eigenstates  using Jacobi coordinates $\boldsymbol{R}$ and $\boldsymbol{r}$ defined in Fig.~\ref{fig:engfile}. That is, 
\begin{equation}
       \langle\boldsymbol{R},\boldsymbol{r}|\Psi\rangle = \sum_i F_{i}(R)\,  \langle\hat {\boldsymbol{R}},\boldsymbol{r} |\psi_i\rangle\,,
       \label{eq:basis1}
\end{equation}
where index $i$ labels basis states or channels $|\psi_i\rangle$. 
The sum  over channels  is constrained to conserve total angular momentum of the trimer $\boldsymbol{F}_{\rm tot}$ and parity $p_{\rm tot}=\pm 1$.

Channel states $|\psi_i\rangle$ are given by 
\begin{eqnarray}
\langle\hat {\boldsymbol{R}},\boldsymbol{r} |\psi_i\rangle  &=& 
\sum_{mM}C^{F_{\rm tot} M_{\rm tot}}_{\ell m,FM}\,Y_{\ell m}(\hat{\boldsymbol{R}})\,\langle \boldsymbol{r} |\varphi_{\tau;FM;p}\rangle,
\label{eq:basis2}
\end{eqnarray}
where $C^{jm}_{j_1m_1,j_2m_2}$ are Clebsch-Gordan coefficients, $Y_{\ell m}(\hat {\boldsymbol{R}})$ are  spherical harmonic functions
of the orbital angular momentum or partial wave $\boldsymbol{\ell}$,
and $|\varphi_{\tau;FM;p}\rangle$ describe electronic  and OH ro-vibrational states with combined angular 
momentum $\boldsymbol{F}$ and parity $p$. Label $\tau$ is used to further describe these states.
Projection quantum numbers  $M_{\rm tot}$, $m$, and $M$ of $\boldsymbol{F}_{\rm tot}$, $\boldsymbol{\ell}$, and  $\boldsymbol{F}$,
respectively, are given with respect to a space-fixed laboratory axis.
Here, $\boldsymbol{F}_{\rm tot} = \boldsymbol{\ell} + \boldsymbol{F} $, parity $p_{\rm tot}=(-1)^\ell p$.

In principle,  states $|\varphi_{\tau;FM;p}\rangle$ form a complete set of electronic  and OH ro-vibrational states.
In order to make the computations tractable, however, we must introduce various approximations.
The first is that the electronic wavefunctions are restricted to be the non-relativistic {\it diabatic} states $|1\,^2\!A^{\prime}({\rm X}^2\Sigma^+)\rangle$, $|4\,^2\!A^{\prime}({\rm F}^2\Pi)\rangle$, and $|4\,^2\!A^{\prime\prime}({\rm F}^2\Pi)\rangle$,  defined in the main text. These 
diabatic states are assumed to be independent of $R$, $r$, and $\theta$ (an approximation already implicit in the notation
used in Eqs.~\ref{eq:basis1} and \ref{eq:basis2}.) We  further assume that these
trimer electronic wavefunctions are given by simple products  of Sr($^1$S) and either $^2\Pi$ and $^2\Sigma^+$ $^{16}$OH 
dimer electronic wavefunctions.
That is, $|4\,^2\!A^{\prime}({\rm F}\,^2\Pi)\rangle$ and $|4\,^2\!A^{\prime\prime}({\rm F}\,^2\Pi)\rangle$
are superpositions of $|{\rm Sr}(^1{\rm S})\rangle | {\rm OH}({\rm X}\,^2\Lambda)\rangle$ with projection quantum number 
${\Lambda=\pm\Pi}$ or $\pm1$ and 
$|1\,^2\!A^{\prime}({\rm X}^2\Sigma^+)\rangle\to |{\rm Sr}(^1{\rm S})\rangle | {\rm OH}(^2\Sigma^+)\rangle$ 
(The state has projection $\Lambda=\Sigma$ or 0),
re-enforcing the usefulness of labeling  trimer electronic states with $C_{\infty{\rm v}}$ symmetries at collinear geometries.

Secondly, the vibrational motion of the OH stretch is limited to that of the ${v=0}$ state of  $^{16}$OH(X$^2\Pi$). This
approximation relies on the observations that the $4\,^2\!A^{\prime}({\rm F}^2\Pi)$ and $4\,^2\!A^{\prime\prime}({\rm F}^2\Pi)$
potentials are less deep than the vibrational spacing of OH(X$^2\Pi$) and that near the conical intersections of
the $|1\,^2\!A^{\prime}({\rm X}^2\Sigma^+)\rangle$ and $|4\,^2\!A^{\prime}({\rm F}^2\Pi)\rangle$ states the $r$ dependence of the
$1\,^2\!A^{\prime}({\rm X}^2\Sigma^+)$ potential, except for a constant energy offset, is nearly independent of $R$.
Near the equilibrium geometry of the  $1\,^2\!A^{\prime}({\rm X}^2\Sigma^+)$ potential the vibrational energy in the OH stretch is within 
$3\,\%$ of that of the OH(X$^2\Pi$) dimer. 

Spin-orbit interactions and OH rotation  are included based on our assumption that trimer electronic states are separable and determined 
by those in the OH dimer. Then spin-orbit interactions in the $|1\,^2\!A^{\prime}({\rm X}^2\Sigma^+)\rangle$  state
are absent while those in the $|4\,^2\!A^{\prime}({\rm F}^2\Pi)\rangle$ and $|4\,^2\!A^{\prime\prime}({\rm F}^2\Pi)\rangle$ states
correspond to those in the OH(X$^2\Pi$) dimer. States represented by $^2\Sigma^+_{\Omega=1/2}$ and $^2\Pi_{\Omega=1/2,3/2}$ with 
OH electronic angular momentum projection $\Omega=1/2$ and $3/2$  on the internuclear axis of $^{16}$OH are formed.
The rotation, coriolis and hyperfine interactions,  as well as $\Lambda$-doubling in SrOH are similarly based on  those in OH.
In fact, these latter interactions weakly mix $^2\Pi_{\Omega=1/2}$ and $^2\Pi_{\Omega=3/2}$ states as well as mix  electron-rotational
angular momentum states $\boldsymbol{J}$.
See Refs.~\cite{Petrov:13,Baklanov2010} for a discussion of symmetries of and the spin-orbit, hyperfine, $\Lambda$-doublet interactions in OH. 

In practice, the relevant states are 
\begin{eqnarray}
\lefteqn{ \langle \boldsymbol{r} |\varphi_{\tau;FM;p}\rangle = \varphi_{v=0,J\Omega}(r) }
\label{eq:basis3}
\\
&&\quad\quad \times\sum_{M_JM_I} C^{F M}_{JM_J,IM_I}  \,\langle \hat {\boldsymbol{r}} | ^2\Lambda_\Omega, JM_J \Omega; p \rangle|    I M_I\rangle \,,
\nonumber
\end{eqnarray}
where radial functions $\varphi_{v=0,J\Omega}(r)$ are   ${v=0,J}$ OH(X$^2\Pi_\Omega$) rotational levels, obtained by numerical diagonalization of the OH Hamiltonian excluding the coriolis, hyperfine, $\Lambda$-doubling interactions \cite{Petrov:13}, and
 OH electron-rotational states  $| ^2\Lambda_\Omega, JM_J \Omega; p \rangle$ are
\begin{eqnarray}
\lefteqn{\langle \hat {\boldsymbol{r}} | ^2\Lambda_\Omega, JM_J \Omega; p \rangle=
|{\rm Sr}(^1{\rm S})\rangle\frac{1}{\sqrt{2}} \Bigl\{ \theta^{J}_{M_J,\Omega}(\hat {\boldsymbol{r}})|{\rm OH}(^2\Lambda_\Omega)\rangle}
\nonumber\\
&&\quad  \quad\quad +\, (-1)^{J+2\Omega+1/2}\,p \,\theta^{J}_{M_J,-\Omega}(\hat {\boldsymbol{r}})
|{\rm OH}(^2\Lambda_{-\Omega})\rangle \Bigr\} 
\end{eqnarray}
with 
\begin{equation}
 \theta^{J}_{M_J,\Omega}(\hat {\boldsymbol{r}})=\sqrt{\frac{2J+1}{4\pi}}D^{J*}_{M_J,\Omega}(\alpha,\beta,\gamma=0)\,.
\end{equation}
Here, operator $\boldsymbol{J}$ is the electron-rotational angular momentum of $^{16}$OH with quantum number
$J$ and projections $M_J$  and $\Omega$ along the laboratory axis and  $^{16}$OH axis $\boldsymbol{r}$, respectively. 
Functions $D^{J}_{M,\Omega}(\alpha,\beta,\gamma)$ are Wigner rotation functions with
Euler angles $\alpha$, $\beta$, and $\gamma$ that specify the orientation of OH and, thus, $\boldsymbol{r}$ in our 
space-fixed coordinate system.  For the $^{16}$OH electronic wavefunctions $|{\rm OH}(^2\Lambda_\Omega)\rangle$
we have $\Lambda=\Sigma$ and $\Pi$ (or 0 and 1).
Finally, the ket $|I M_I\rangle $ describes the nuclear spin wavefunction of H with $M_I=\pm1/2$
and total $^{16}$OH angular momentum $\boldsymbol{F} = \boldsymbol{J} + \boldsymbol{I} $. Consistent with our
approximations we have ignored the  electronic state dependence of $\varphi_{v=0,J\Omega}(r)$.
Historically, $e$ states are  states with $p=(-1)^{J-1/2}$ while  $f$ states are  states with $p=(-1)^{J+1/2}$.
In summary, index $i$ in Eq.~\ref{eq:basis1} represents labels and quantum numbers $^{2}\Lambda_\Omega; p, [(JI)F, \ell ]F_{\rm tot} M_{\rm tot};p_{\rm tot}$. Similarly,  $\tau$ in Eq.~\ref{eq:basis2} represents $^{2}\Lambda_\Omega;  JI$. Reference \cite{Alexander1982} gives details of the required angular momentum algebra for a simpler case.

Finally, the radial wave functions $F_{i}(R)$ are  channel components and only depend on the  separation  $R$ between
the center of mass of the molecule and the atom. They are determined from numerically integrating the coupled radial Schr\"odinger equations
\begin{equation}
	\left\{-\frac{\hbar^2}{2\mu}\frac{d^2}{dR^2} + \boldsymbol{U}(R) \right\}\boldsymbol{F}(R) = E\,\boldsymbol{F}(R)
\end{equation}
for total energy $E$ from $R=0$ to $R\to\infty$. Here, vector $\boldsymbol{F}(R)$  contains all $F_i(R)$ and  $\mu$ is the reduced mass of the atom-molecule system. The first term in curly brackets on the left hand side of this equation is a diagonal matrix for
the radial kinetic energy (the same for each channel), while the second term $\boldsymbol{U}( R) $ is the potential  matrix with  elements $U_{ij}(R)=\langle\psi_i|U|\psi_j\rangle$ between
channels $i$ and $j$ and includes the three {\it diabatic} potentials $V(R,r,\theta)$ and their couplings, the rotation energy
of Sr around OH, $\hbar^2\boldsymbol{\ell}/(2\mu R^2)$, as well as the spin-orbit, rotational
and hyperfine energies of the $v=0$ vibrational level of OH.   In fact, $\boldsymbol{U}(R)$ is anisotropic, orientation dependent 
as channels with different values for $\ell$ or $m$ are coupled.  
We limit the number of channel components by only including $^{16}$OH rotational states $J\le 25/2$.

At large separations $\boldsymbol{U}( R)$ is an ``almost''  diagonal matrix. For the basis in Eq.~\ref{eq:basis3} weak couplings by OH  coriolis, hyperfine and $\Lambda$-doubling interactions persist and mix states with different values of $\Omega$ and $J$. Hence, in order to compute scattering amplitudes and rate coefficients we first calculate the orthonormal matrix that diagonalizes 
$\boldsymbol{U}(R)$ at a large $R$ and transform $\boldsymbol{F}(R)$ into the basis in which $\boldsymbol{U}( R)$ is diagonal.

\section{Acknowledgment}
Work at Temple University is supported by the Army Research Office Grant No. W911NF-17-1-0563 and the National Science Foundation Grants Nos. PHY-1619788 and PHY-1908634.

\vspace{0.5cm}
\section{Author Contributions}
S.K. supervised the project, M.L. and A.P.  developed and coded the numerical methods for the scattering calculations. M.L, J.K., and H.L performed {\it ab~initio} and scattering
calculations, S.K.  wrote first draft of the manuscript with contributions from M.L, J.K., and A.P.

\section{Additional Information}
{\bf Competing financial interests:} The authors declare no competing financial interests. 

{\bf Data availability}: All data generated or analysed during this study are available upon request.

\bibliographystyle{naturemag}
\bibliography{refs_new_10}

\end{document}